\documentclass[12pt]{article}
\usepackage{amsmath,amssymb,bm,epsfig,afterpage}
\usepackage{cite}
\usepackage{here}
\usepackage{color}

\renewcommand{\thefootnote}{\fnsymbol{footnote}}

\newcommand{\MEW}{m_Z}

\begin{document}
\title{
\begin{flushright}
\begin{minipage}{0.2\linewidth}
\normalsize
WU-HEP-16-02 \\*[50pt]
\end{minipage}
\end{flushright}
{\Large \bf 
Constraints on non-universal gaugino mass scenario using the latest LHC data
\\*[20pt]}}
\author{
Junichiro~Kawamura$^a$\footnote{
E-mail address: junichiro-k@ruri.waseda.jp} \ and
Yuji~Omura$^b$\footnote{
E-mail address: yujiomur@eken.phys.nagoya-u.ac.jp}\\*[20pt]
{\it \normalsize 
$^a$Department of Physics, Waseda University, 
Tokyo 169-8555, Japan} \\
{\it \normalsize 
$^b$Kobayashi-Maskawa Institute for the Origin of Particles and the Universe (KMI), } \\
{\it \normalsize
Nagoya University, Nagoya 464-8602, Japan} \\*[50pt]
}
\date{
\centerline{\small \bf Abstract}
\begin{minipage}{0.9\linewidth}
\medskip 
\medskip 
\small
We investigate exclusion limits on the the non-universal gaugino mass scenario 
in the Minimal Supersymmetric Standard Model (MSSM), according the the latest results of the super-particle search at the LHC8 and the LHC13. In this scenario, suitable ratios of wino to gluino mass can realize the observed value of the Higgs boson mass, while keeping a small $\mu$ parameter. 
Such a small $\mu$ parameter corresponds to the mass of higgsino, so that 
lightest neutralino and chargino are higgsino-like and their masses are almost degenerate.
Besides, we find that the right-handed top squark tends to be lighter than other sfermions 
and then the top squark search, where the top squark decays to a quark and higgsino, is relevant to our model.
In our analysis, the exclusion limits are derived using the data of the top squark searches 
in the $bb + E_T^{\rm miss}$ and $tb + E_T^{\rm miss}$ channels.
Furthermore, the exclusion limit on gluino mass, which is crucial to our scenario, is investigated as well.
The analysis of the gluino is based on the data of the analysis with large missing energy and at least three b-tagged jets at the ATLAS experiment.  
\end{minipage}
}

\begin{titlepage}
\maketitle
\thispagestyle{empty}
\clearpage
\tableofcontents
\thispagestyle{empty}
\end{titlepage}

\renewcommand{\thefootnote}{\arabic{footnote}}
\setcounter{footnote}{0}

\section{Introduction}

The Higgs boson discovery is one of the most significant results 
at the Large Hadron Collider (LHC)~\cite{Aad:2012tfa,Chatrchyan:2012ufa},
and giving a good insight to new physics beyond the Standard Model (SM).
For instance, supersymmetry (SUSY) is one of the promising candidates for 
the new physics, and the LHC surveys the signals of SUSY particles.
However, we know that there is a tension between  
the observed Higgs boson mass and the prediction of of the Minimal Supersymmetric Standard Model (MSSM)~\cite{Martin:1997ns}.
This problem is called the little hierarchy problem, and has been widely discussed
even before the LHC started.

This problem is concerned with the so-called $\mu$ parameter, which corresponds to the supersymmetric masses of Higgs bosons and higgsinos.
In the MSSM, the electro-weak (EW) symmetry breaking is realized by nonzero vacuum expectation values (VEVs)
of the two Higgs doublets. Then, the stationary conditions for the EW symmetry breaking predict the following relation
between the EW scale and the SUSY breaking scale:
\begin{eqnarray}
m_Z^2 &\simeq -2 \left| \mu(\MEW) \right|^2 -2 m_{H_u}^2(\MEW), 
\label{EWSB}
\end{eqnarray}      
where $\mu(\MEW)$ and $m_{H_u}(\MEW)$ 
are the $\mu$ parameter and a soft SUSY breaking mass for the up-type Higgs boson 
at the Z boson mass scale  ($m_Z \simeq 91.2$ GeV), respectively.   
We expect that soft SUSY breaking parameters are generated dynamically, according to spontaneous SUSY breaking and a certain mediation mechanism. Then, all soft SUSY breaking parameters are expected to be at the same scale.
On the other hand, the Higgs mass is predicted to be less than $m_Z$ at the tree-level in the MSSM. 
The loop corrections involve top squark masses mainly, and then 125 GeV Higgs mass can be achieved,
if the top squarks are more than ${\cal O}(1)$ TeV,
as long as the trilinear coupling of the top squarks (A-term) does not satisfy a certain relation
(so-called maximal mixing). 
Eventually, such a high SUSY breaking scale tends to predict large $m_{H_u}(\MEW)$,
so that we have to realize the miracle cancellation between
the supersymmetric mass $\mu$ and the SUSY breaking scales to satisfy Eq. (\ref{EWSB}).

One simple solution of the little hierarchy problem has been proposed in Refs.~\cite{Abe:2007kf,Abe:2012xm}.  
A key ingredient is a suitable ratio of wino to gluino mass at the gauge coupling unification scale. 
If the wino mass is about 5 times heavier than the guino, $m_{H_u}(\MEW)$ becomes
EW-scale and the maximal mixing is easily realized. Then we do not need require the miracle cancellation
and $125$ GeV Higgs mass is also realized within the framework of the MSSM.  
Let us call this scenario as Non-Universal Gaugino Mass (NUGM) scenario.

It is known that such a specific relation between gaugino masses can be realized 
by suitable structure of gauge kinetic function at the tree level 
in string models~\cite{Abe:2005rx,Blumenhagen:2006ci,Sumita:2015tba} and the Grand Unified Theory (GUT)~\cite{Younkin:2012ui}. Furthermore, we find that the mirage mediation which is a mixture of the moduli 
and anomaly mediation can realize the NUGM scenario~\cite{Randall:1998uk}. 
Note that phenomenologies of the mirage mediation have been also studied 
in Refs.\cite{Endo:2005uy,Choi:2005uz,Choi:2006xb,Kitano:2005wc,Abe:2014kla}. 

In the NUGM scenario, the lightest supersymmetric particle (LSP) is higgsino-like,
corresponding to the small $\mu$. 
The right-handed top squark also tends to be light, 
because of the heavy wino, as we discuss in Sec. \ref{sec2}.  
In this paper, we extract exclusion limits on the parameter space of the NUGM scenario 
based on the latest results at the LHC, especially focusing on the direct searches for top squarks and gluinos.
In Ref. \cite{Abe:2015xva}, the authors discussed the properties of the top squark decays, 
and drew exclusion limits on the top squark mass with the fixed $\mu$ parameter  
based on the experimental results in the $bb +  E_T^{\rm miss}$ channel at the ATLAS.
Here, we study the constraints coming from both gluino and top squark searches 
referring the latest data at the LHC13~\cite{LHC13:sb,LHC13:glu}
 in addition to the ATLAS analyses at the LHC8~\cite{Aad:2013ija,Aad:2015pfx,Aad:2014lra}.
Moreover, we will survey more wider region than the work in Ref. \cite{Abe:2015xva}, varying $\mu$ parameter and 
considering large $\tan\beta$ case.

We note that  
it might be difficult to explain whole abundance of the dark matter 
by the thermally produced neutralino LSP~\cite{Mizuta:1992qp}, since the LSP is the higgsino-like neutralino.
Some extensions of the model or new cosmological scenario would be necessary, 
e.g. introducing axions~\cite{Rajagopal:1990yx}, axinos~\cite{Covi:1999ty}, singlinos 
in the Next MSSM~\cite{Ellwanger:2009dp} and/or non-thermal production of the LSP~\cite{Kohri:2005ru},  
although this is beyond the scope of this paper.

This paper is organized as follows. 
In Section 2, we explain how the suitable ratio of wino to gluino mass parameter 
can enhance the Higgs boson mass and solve the little hierarchy problem. 
The typical mass spectrums are shown in Subsection 2.2. 
In Section 3, the current exclusion limits on the NUGM scenario are exhibited.   
We explain our strategy to extract exclusion limits, 
and then the exclusion lines from the top squark and gluino searches are shown in Subsection 3.1 and 3.2, respectively. 
In Subsection 3.3, we will discuss possibility of the additional contribution to the top squark searches 
from the bottom squark production in the large $\tan\beta$ case.
We conclude this paper in Section 4.

\section{Non-Universal Gaugino Masses scenario}
\label{sec2}
\subsection{Overview}

First, let us shortly review the NUGM scenario.
The Higgs boson mass in the MSSM can be written approximately as~\cite{Carena:1995wu}, 
\begin{eqnarray} 
m_h^2 \simeq 
m_Z^2 \cos^2 2 \beta+\frac{3}{8 \pi^2} \frac{m_t^4}{v^2} \left[\log{\frac{M_{\rm st}^2}{m_t^2}}+ 
 \frac{2\widetilde{A}_t^2}{M_{\rm st}^2} 
\left( 1-\frac{\widetilde{A}_t^2}{12 M_{\rm st}^2} \right)  \right], 
\label{Mhig}
\end{eqnarray}
where $\tilde{A}_t \equiv A_t - \mu \cot\beta$ is defined.
The symbols $m_t$, $M_{\rm st}\equiv \sqrt{|m_{Q_3}m_{u_3}|}$ and $A_t$ denote 
the top quark mass, the top squark mass scale 
and the left-right mixing of the top squarks, called  A-term, respectively.  
Here, $m_{u_3}$ and $m_{Q_3}$ are the soft scalar masses for the right-handed and left-handed third-generation squark. 
It is known that the Higgs boson mass requires $M_{\rm st} \simeq 10$ TeV 
unless $A_t / M_{\rm st} \simeq \sqrt{6}$, the so-called maximal mixing scenario, is satisfied. 

Next, let us discuss the Renormalization Group (RG) runnings of the soft SUSY breaking terms.
The values of $m_{H_u}, m_{Q_3}, m_{u_3}$ and $A_t$ at the EW scale 
depend on the boundary condition at the GUT scale as follows~\cite{Abe:2012xm}: 
\begin{align}   
m_{H_u}^2(\MEW) &\simeq - 0.01 M_1 M_2 + 0.17 M_2^2 - 0.05 M_1 M_3 - 0.20 M_2 M_3 - 3.09 M_3^2 \notag \\
&+ (0.02 M_1 + 0.06 M_2 + 0.27 M_3 - 0.07 A_t) A_t + 0.59m_{H_u}^2 - 0.41m_{Q_3}^2 - 0.41m_{U_3}^2, \label{RGE_mhu}  \\
m_{Q_3}^2(\MEW) &\simeq -0.02 M_1^2+0.38 M_2^2-0.02 M_1 M_3-0.07 M_2 M_3+5.63 M_3^2 \notag \\ 
&+(0.02 M_2+0.09 M_3-0.02 A_t)A_t-0.14m_{H_u}^2+0.86m_{Q_3}^2-0.14 m_{U_3}^2, \label{RGE_Q} \\ 
m_{u_3}^2(\MEW) &\simeq 0.07 M_1^2-0.01 M_1 M_2-0.21 M_2^2-0.03 M_1 M_3-0.14 M_2 M_3+4.61 M_3^2 \notag \\ 
&+(0.01M_1+0.04 M_2+0.18 M_3-0.05 A_t) A_t-0.27 m_{H_u}^2-0.27 m_{Q_3}^2+0.73m_{U_3}^2, \label{RGE_u} \\
A_t(\MEW) &\simeq -0.04 M_1-0.21 M_2-1.90 M_3+0.18 A_t, \label{RGE_At}
\end{align} 
where $M_i\ (i=1,2,3)$ are the gaugino mass parameters at the GUT scale.
The other parameters in the right-hand side are the ones at the GUT scale as well.
We can see that the RG effects are dominated by the gluino mass parameter $M_3$, 
especially the ratio $A_t / M_{\rm st}$ at $\MEW$ is less than unity as long as the gluino mass dominates the RG effects.

This situation is altered, if the wino mass is much heavier than the gluino mass at the GUT scale.
Eq.(\ref{RGE_u}) tells that the right-handed top squark mass $m_{u_3}$ decreases while the $A_t$ increases 
as the wino mass parameter increases. 
Thus the ratio $A_t/M_{\rm st}$ can excess unity and could reach $A_t/M_{\rm st} = \sqrt{6}$ at $\MEW$. 
Note that there is an upper bound on the ratio of wino to gluino mass 
in order to avoid the tachyonic right-handed top squark, $m_{u_3}^2 < 0$, that is, $M_2/M_3 \lesssim 5$. 
In other words, the ratio $A_t / M_{\rm st}$ is maximized at $M_2/M_3 \simeq 5$.

It is interesting that the absolute value of $m_{H_u}$ tend to be small as the wino mass parameter increases. 
The RG contributions from the wino mass parameter can cancel those from the gluino mass parameter, 
because the coefficient of the $M_2^2$ term is positive while it is negative for the $M_3^2$ term. 
Furthermore, the gluino contributions are canceled almost completely by the wino contributions in Eq.(\ref{RGE_mhu}) when the ratio satisfies $M_2/M_3 \simeq 5$. 
Therefore, the value of the $\mu$ parameter can be small and the Higgs boson mass is enhanced 
around the region with $M_2/M_3 \simeq 5$, where $m_{u_3}^2$ becomes small as well.

\subsection{Sparticle spectrum}
The relatively large wino mass induces specific characters of sparticle spectrums.
Clearly, the left-handed sparticles 
tend to be heavier than the right-handed ones, because of the RG running involving $M_2$.
In fact, the right-handed top squark mass $m_{u_3}$ is suppressed, 
since the large left-handed top squark mass $m_{Q_3}$ reduces $m_{u_3}$ through the top Yukawa coupling 
in the RG effects. 
Note that this is one of the reasons why the Higgs boson mass can reach the observed value 
even when the top squark is less than sub TeV. 
It is also specific to large wino case that sleptons are as heavy as squarks.

\begin{table}[!htb]
\centering
\begin{tabular}{|c|c|c|c|c|c|c|}\hline
  input [GeV]& ST1 & ST2 & SB1 & SB2 & GL1 & GL2 \\ \hline\hline
 $\mu(\MEW)$ & 150 & 300 & 150 & 150 & 150 & 700 \\
 $\tan\beta$   & 15 & 15 & 50 & 50 & 15 & 15 \\
 $A_0$ & -1800 & -1570 & -2250 & -2710 & -3300 & -3530 \\
 $M_1$ & 5500 & 4000    &  5500  & 8000   & 12000 & 12000 \\
 $M_2$ & 4365 & 4183    &  4321  & 4153   & 4294  &  3867   \\
 $M_3$ & 1000 & 1000    &  1000  & 800   & 550 &  400     \\ \hline\hline
 mass [GeV] & & & & & & \\ \hline\hline
 $m_h$                      & 126.1& 125.8 & 126.0  &126.1 & 126.0 & 125.8 \\
 $m_H$                     & 2810  & 2633  & 906.1 & 981.3 & 3337 & 3197  \\
 $m_{\widetilde{d}_L}$ & 3427 &  3329 & 3415 & 3181 & 3117 & 2807 \\
 $m_{\widetilde{d}_R}$ & 2175 & 2129 &  2184 & 2020 & 2004 & 1883 \\
 $m_{\widetilde{u}_L}$ & 3426 & 3328 & 3414 & 3180 & 3116 & 2806 \\
 $m_{\widetilde{u}_R}$ & 2458 & 2286 & 2467 & 2623 & 3224 & 3152 \\
 $m_{\widetilde{b}_1}$ & 2077 & 2039 & 866.2 & 372.4 & 1863 & 1738 \\
 $m_{\widetilde{b}_2}$ & 2960 & 2910 & 2629 & 2272   & 2314 & 1958 \\
 $m_{\widetilde{t}_1}$ & 587.4 & 449.2 & 657.1 & 934.1 & 1489 & 1506 \\ 
 $m_{\widetilde{t}_2}$ & 2968 &2917   & 2638 & 2286 & 2334 & 1996  \\
 $m_{\widetilde{e}_L}$ & 3107 & 2921 & 3086 & 3182 & 3642 & 3441  \\
 $m_{\widetilde{e}_R}$ & 2256 & 1778 & 2258 & 3109 & 4522 & 4520  \\
 $m_{\widetilde{\tau}_1}$ & 2186 & 1706 & 1353 & 2307 & 3596 & 3392  \\
 $m_{\widetilde{\tau}_2}$ & 3082 & 2900 & 2809 & 2820 & 4446  & 4445  \\
 $m_{\widetilde{g}}$                 &2236 & 2230 & 2235 & 1835 & 1317 & 986.2  \\
 $m_{\widetilde{\chi}^{\pm}_1}$ & 154.7 & 305.4 & 155.3 & 154.1 & 155.2 & 712.3 \\
 $m_{\widetilde{\chi}^{\pm}_2}$ & 3545 & 3400 & 3519 & 3379  & 3488 & 3147  \\
 $m_{\widetilde{\chi}^0_1}$      & 152.9 & 303.5 & 153.5 & 152.3 & 153.8 & 710.7 \\
 $m_{\widetilde{\chi}^0_2}$      & 156.2 & 307.5 & 156.8 & 155.4 & 156.3  & 713.8 \\
 $m_{\widetilde{\chi}^0_3}$      & 2442 & 1776 & 2446 & 3379    & 3488 & 3146  \\
 $m_{\widetilde{\chi}^0_4}$      & 3545 & 3400 & 3519 & 3560    & 5348 & 5364  \\ \hline\hline

\end{tabular}
\caption{Values of input parameters, sparticle masses and Higgs boson masses at several sample points. 
A value of universal soft mass $m_0$ is fixed at 1~TeV. 
}
\label{tab_spec}
\end{table}

The Table~\ref{tab_spec} exhibits the values of input parameters. The sparticle masses at several sample points are calculated 
using the SOFTSUSY~\cite{Allanach:2001kg}.  For simplicity, we assume the universal values for the A-terms $A_0$ and soft scalar masses $m_0$ at the GUT scale, while not for gaugino masses through this paper. 
The branching fractions of the top squark, sbottom decays and the gluino decay are shown in 
Table~\ref{tab_st} and Table~\ref{tab_gl}, respectively. 
The branching fractions are calculated using the SDECAY~\cite{Muhlleitner:2003vg}.

In the NUGM, the LSP is higgsino-like. 
The masses of higgsino-like states, 
namely the lightest and the second lightest neutralino $\tilde{\chi}_{1,2}^0$ 
and the lightest chargino $\tilde{\chi}_{1}^\pm$, 
are nearly degenerate. 
Typical mass differences between the neutralinos and the chargino is about a few GeV  
as shown in Ref.~\cite{Abe:2015xva}.
In this case, signals from the higgsinos are quite difficult to be distinguished from the SM backgrounds 
due to the small mass difference among the higgsino-like states.  
Besides, the chargino could not be detected as disappeared tracks unlike the wino LSP 
case~ \cite{Aad:2013yna,CMS:2014gxa}
that the mass difference is typically a few handred MeV. 
Thus, the higgsino will be quite hard to be detected directly at the LHC, 
although there are several works to search for nearly degenerate 
higgsinos~\cite{Han:2013usa,Baer:2013xua,Baer:2014kya,Baer:2013yha}.

\begin{table}[!tbh]
\centering
\begin{tabular}{|c|c|c|c|c|c|c|c|}\hline
  branching ratio& ST1 & ST2 & SB1 & SB2 & branching ratio &SB1 & SB2 \\ \hline\hline
 Br$(\widetilde{t}_1 \rightarrow t \widetilde{\chi}^0_{1,2})$    & 0.451 & 0.000   & 0.463 & 0.484 & Br$(\widetilde{b}_1 \rightarrow t \widetilde{\chi}^{\pm}_1)$ & 0.480 & 0.299 \\  
 Br$(\widetilde{t}_1 \rightarrow b \widetilde{\chi}^{\pm}_1)$ & 0.549 & 1.000 & 0.537 & 0.516  &  Br$(\widetilde{b}_1 \rightarrow b \widetilde{\chi}^0_{1,2}) $   &0.520 & 0.701 \\ \hline
\end{tabular}
\caption{Branching ratios of the top squark decay at the sample points.}
\label{tab_st}
\end{table}

One powerful way to probe the NUGM, instead of the direct search for higgsino, is the top squark search.  
Since our top squark is almost right-handed and the masses of the higgsinos are nearly degenerate, 
the top squark can decay the third-generation quark and the higgsino-like particle with branching fractions:  
\begin{eqnarray}
{\rm Br}(\tilde{t}_1 \rightarrow t \tilde{\chi}_{1,2}^0) \simeq {\rm Br}(\tilde{t}_1\rightarrow b \tilde{\chi}_1^\pm) \simeq 0.5  
\end{eqnarray}
as long as the decay $\tilde{t}_1 \rightarrow t \tilde{\chi}_{1,2}^0$ is kinematically unsuppressed. 
The degeneracy of the higgsinos also makes the 
decays of the second lightest neutralino and the lightest chargino invisible, 
and then a pair produced top squarks will be detected as $tt\ {\rm or}\ tb\ {\rm or}\ bb + E_T^{\rm miss}$. 
It is known that the search for $bb + E_T^{\rm miss}$ channel gives more stringent bound 
than those for $tt + E_T^{\rm miss}$ channel 
when the top squark is right-handed one 
and enough light to be tested by the current data of the LHC~\cite{Abe:2015xva,Han:2013kga}.

Similarly, the right-handed bottom squark tends to be lighter than other sparticles 
through the bottom Yukawa coupling in the RG running 
when the bottom Yukawa coupling is as large as the top Yukawa coupling.
This situation corresponds to large $\tan\beta$.
The right-handed bottom squark will decay into $b + \tilde{\chi}^0_{1,2}$ or $t + \tilde{\chi}^\pm_1$, 
with each branching ratio is about 0.5. 
Thus the bottom squark predicts same signal as the top squark, 
and then the exclusion limits would be tightened by such contributions.

The third and fourth columns of Table.~\ref{tab_st} show the sparticles masses with $\tan\beta=50$. 
At the sample point SB1, 
the bottom squark mass is heavier than the top squark about 200 GeV, 
then this will not give sizable contribution to the top squark search. 
On the other hand, the bottom squark is so light at the sample point SB2 
that this point will be excluded by the third generation squark searches.

The mass hierarchy between the top and bottom squarks is determined by details of the RG equations. 
For the top squark, the RG contribution from the gluino mass is almost canceled by those from the wino mass 
as long as the small $\mu$ parameter is realized in the NUGM scenario.
Then the top squark mass is mainly determined by the bino mass parameter.  
On the other hand, such cancellation does not occur for the bottom squark efficiently compared with the top squark mass, and then the gluino mass contribution is also important for the bottom squark mass.

\begin{table}[!tbh]
\centering
\begin{tabular}{|c|c|c|c|c|c|}\hline
  branching ratio & GL1 & GL2 & branching ratio & GL1 & GL2 \\ \hline\hline 
 Br$(\widetilde{g} \rightarrow t t \tilde{\chi}^0_{1,2})$    & 0.438 & 0.000 & Br$(\widetilde{g} \rightarrow b b \tilde{\chi}^0_{1,2})$  &0.015 & 0.096 \\
 Br$(\widetilde{g} \rightarrow t b \tilde{\chi}^{\pm}_1)$ & 0.535 & 0.253 &Br$(\widetilde{g} \rightarrow g \tilde{\chi}^0_{1,2})$   &0.012 & 0.651  \\ \hline\hline
\end{tabular}
\caption{Branching fractions of a gluino at the sample points.}
\label{tab_gl}
\end{table}

The other important observation to test the NUGM scenario is the gluino search. 
The bino and wino mass parameter is typically larger than the gluino mass parameter at the GUT scale, 
then their contributions push up the masses of the sleptons 
compared with the universal gaugino mass case. 
As usual, the squark masses are raised by the gluino mass contributions. 
These facts lead that all sparticles except top (bottom) squark and higgsino have 
same or larger masses than gluino as we can see in Table.\ref{tab_spec}. 
The gluino has the largest production cross section at the LHC if the sparticle masses are the same-scale, 
thus gluino searches will also give the stringent bound on the NUGM scenario. 
The typical mass spectrums at the sample points GL1, GL2 
with the light gluino are exhibited in the fifth and sixth columns 
of the Table.~\ref{tab_spec}.

The branching fractions of the gluino decay at the sample points GL1, GL2 are exhibited in the Table~\ref{tab_gl}. 
The gluino decays into on/off-shell third-generation squarks 
when the mass difference between the gluino and the higgsinos are large enough, 
while a branching fraction of loop induced two-body decay $\tilde{g} \rightarrow g \tilde{\chi}_{1,2}^0$ 
becomes sizable as the mass difference decreases. 
This two-body decay is mediated by third-generation squarks.

\section{LHC constraints on NUGM}
In this section, we draw exclusion limits on top squark and gluino masses 
with varying higgsino mass parameter, using the data at the LHC Run I 
and Run II. 

Our parameter setting is as follows.
We fix $\tan \beta$ at $\tan\beta = 15$ except for the analysis in Subsection 3.3.
The sign of the $\mu$ parameter is positive 
and the universal soft scalar mass $m_0 = 1$ TeV is adopted. 
$A_0$ and $M_2$ are chosen  
to realize the Higgs boson mass in a range $125.5 \le m_h \le 126.1$ GeV 
and a given value of the $\mu$ parameter. 
Note that there is no solution of $A_0$ and $M_2$ to satisfy the above requirements 
and stability of the realistic EW symmetry breaking minimum of the MSSM scalar potential 
if the both $M_1$ and $M_3$ are too small.
Especially, top squark tends to be tachyonic, in this case.

We employ MadGraph5~\cite{Alwall:2014hca} to simulate signal events 
with a pair produced top squarks or gluinos up to one additional parton,  
and then they are passed into PYTHIA6~\cite{Sjostrand:2006za} and DELPHES3~\cite{deFavereau:2013fsa} 
to carry out parton showering and fast detector simulation. 
The matrix element is matched with parton shower using the MLM scheme~\cite{Caravaglios:1998yr},  
and generated hadrons are clustered using the anti-$k_T$ algorithm~\cite{Cacciari:2008gp} with the radius parameter $\Delta R = 0.4$. 
The simulated events are normalized to production cross sections 
of a pair of top squarks and gluinos calculated by the PROSPINO2.1~\cite{Beenakker:1996ed,Kramer:2012bx}. 
We assume that the b-tagging efficiency is 70 \% when we use the data of the LHC8, 
but it is asumed to be 60 \% for the $bb + E_T^{\rm miss}$ channel.  
In the analysis at the LHC13, it is assumed to be 77 \% and 85 \% 
for the top (bottom) squark search and the gluino search, respectively.

\subsection{Top squark search}
As mentioned above, top squark can be lighter than other sparticles, 
and it decays into one third-generation quark and higgsino counted as missing transverse energy. 
The expected signals are $tt\ {\rm or}\ tb\  {\rm or}\ bb + E_T^{\rm miss}$. 
In Ref. \cite{Abe:2015xva}, we have drown the exclusion limits using the data obtained by the ATLAS collaboration\cite{Aad:2014kra, Aad:2013ija}.
The study based on Ref. \cite{Aad:2014kra} is done in the $1l + 4j + E_T^{\rm miss}$ channel aiming to signals from the $tt + E_T^{\rm miss}$, 
while the work on Ref. \cite{Aad:2013ija} analyzes the $bb + E_T^{\rm miss}$ channel. 
The latter gives more stringent bound on top squark mass, so that we use the  $bb + E_T^{\rm miss}$ channel
for the exclusion at the LHC Run I. This result agrees with the one in Ref.~\cite{Han:2013kga}.

A search for $tb + E_T^{\rm miss}$ is an interesting possibility to search for the top squark in the NUGM scenario, 
since the half of a pair produced top squarks will be counted in this channel 
if the mass differences between the top squark and the higgsinos are large enough.
The analysis for $tb + E_T^{\rm miss}$ channel have been done by the ATLAS collaboration~\cite{Aad:2015pfx}.

\begin{figure}[!tb]
\begin{center}
\centering
\hfill
\includegraphics[width=0.8\linewidth]{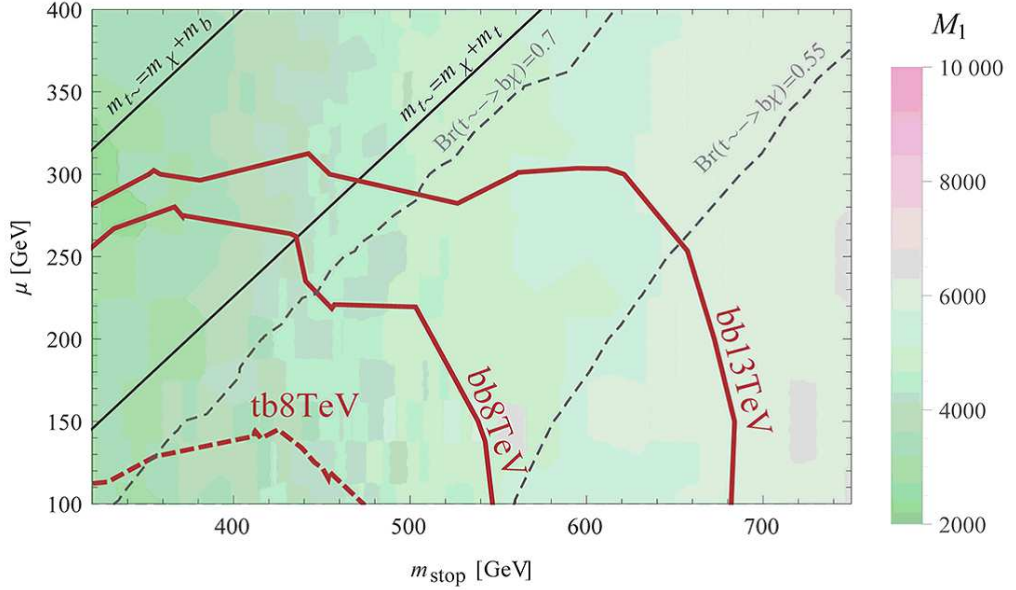} 
\hfill 
\caption{Exclusion limits on the top squark mass with $M_3 = 1.0$ TeV. 
The red solid (dashed) lines depict exclusion limits from the $bb\ (tb)\ + E_T^{\rm miss}$ search. 
The gray dashed lines show the branching fraction of the top squark to a bottom quark and a chargino 
${\rm Br}(\tilde{t}_1 \rightarrow b \tilde{\chi}^{\pm}_1)$. 
The background colors represent the values of bino mass parameter $M_1$ at the GUT scale in the unit of GeV.
}
\label{fig_stmu}
\end{center}
\end{figure}

Fig.\ref{fig_stmu} shows the exclusion limits on the top squark mass and the $\mu$ parameter plane 
with $M_3 = 1.0$ TeV.
The red solid (dashed) lines show the lower bounds using the data of the ATLAS result
dedicated into the $bb\ (tb) + E_T^{\rm miss}$ channel.
To depict these lines, 
we refer the 95$\%$ CL exclusion limits on the number of signal events in each signal region 
exhibited in Table.7 of Ref.\cite{Aad:2013ija}, Table.5 of Ref.\cite{LHC13:sb} for $bb + E_T^{\rm miss}$ channel  
and Table.10 of Ref.\cite{Aad:2015pfx} for $tb + E_T^{\rm miss}$ channel. 
We can see that the $bb + E_T^{\rm miss}$ channel gives the stronger bound than 
those from the $tb + E_T^{\rm miss}$ channel. 
The produced top quarks coming from decays of the top squark cannot be so hard that 
this channel gives more stringent bound than the $bb+E_T^{\rm miss}$ channel, 
because the mass difference between the top squark and the higgsino-like neutralinos is not so large 
at $m_{\tilde{t}} \sim 500$ GeV.  
Furthermore, such a small mass difference also reduces the branching fraction 
${\rm Br} (\tilde{t}_1 \rightarrow t \tilde{\chi}^0_{1,2})$ as we can see from the gray lines of Fig.\ref{fig_stmu},  
especially it almost vanishing when $m_{\tilde{t}_1} \simeq \mu$.  
Thus the $bb + E_T^{\rm miss}$ search is more efficient to probe the NUGM 
at the LHC Run I.

Next, we discuss the exclusion limit according to the latest results at the LHC Run II. 
The lower bound on top squark seems to reach about $700$ GeV for $\mu \lesssim 150$ GeV, 
and goes down to about $600$ GeV at $\mu \simeq 300$ GeV. 
Thus the exclusion limits from the data of the LHC Run II fully cover those from the LHC Run I. 
Furthermore the limit is even sever than the one given by the inclusive razor search\footnote{The razor search excludes the top squark lighter than about 700 GeV, and this result is 
almost independent of the decay patterns of the top squark as long as the LSP mass is enough light, 
although the limit is weakened significantly as the LSP mass increases.}  
at the LHC Run I~\cite{Khachatryan:2015pwa},
although the inclusive search will be suitable to search for top squark in the NUGM since it can cover multiple decay modes.

Let us comment on the difference between the exclusion limits from the $bb + E_T^{\rm miss}$ channel in this paper and the analysis in Ref. \cite{Abe:2015xva}. 
In Ref. \cite{Abe:2015xva}, we simulate the signal events at parton level 
and we do not include the effects of parton showers. 
Specifically, radiation of additional jets will reduce the number of signal events 
in the signal region of $bb + E_T^{\rm miss}$ channel 
where the event with the third high$-p_T$ jet in addition to the leading two b-tagged jets is rejected.

\subsection{Gluino search}
In the NUGM scenario, gluino decays into on/off-shell right-handed top squark:
$\tilde{g} \rightarrow t \tilde{t}_1\rightarrow t t \tilde{\chi}^0_{1,2}\ {\rm or}\ t b \tilde{\chi}^\pm_1$. 
The signal of pair-produced gluinos are characterized by four b-jets 
and large missing energy.  
We refer the upper bound on the number of signal events given 
by the ATLAS collaboration~\cite{Aad:2014lra,LHC13:glu} 
as the top squark search. 
The strategy to calculate the exclusion limits is same as in the previous subsection.

\begin{figure}[!tb]
\centering
\includegraphics[width=0.8\linewidth]{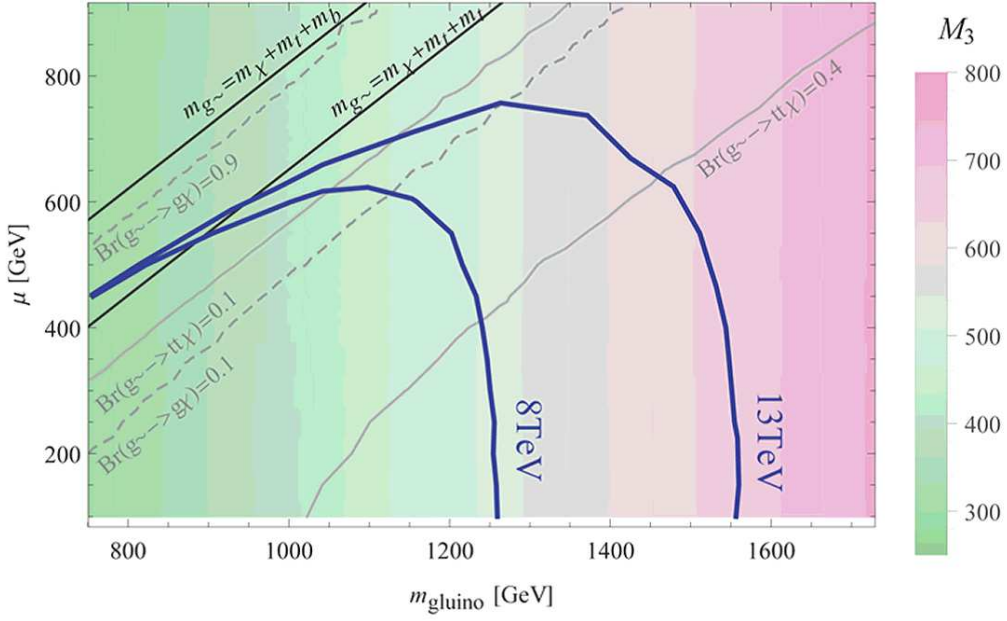} 
\caption{Exclusion limits from the gluino searches with $M_1=12$ TeV are represented by the blue lines 
using the data obtained by the ATLAS~\cite{Aad:2014lra,LHC13:glu}. 
The gray dashed lines are branching fraction of the loop-induced gluino two-body decay 
${\rm Br} (\tilde{g}\rightarrow g \tilde{\chi}_{1,2}^0)$ 
and the gray solid lines are values of branching fraction ${\rm Br} (\tilde{g}\rightarrow t t \tilde{\chi}_{1,2}^0)$. 
The rest of decay modes are dominated by $\tilde{g}\rightarrow t b \tilde{\chi}_{1}^\pm$. 
The background colors represent values of the gluino mass parameter $M_3$ at the GUT scale.}
\label{fig_glmu}
\end{figure}

Fig.\ref{fig_glmu} shows the lower bound on the gluino mass and the $\mu$ parameter plane 
when $M_1$ is fixed at 12 TeV. 
The blue lines are based on the ATLAS results on the gluino search at the LHC Run I and Run II.
The colors on Fig. \ref{fig_glmu}  represent the size of $M_3$. 
As we see, the bound from the LHC Run II already exceed those from the LHC Run I, 
although the integrated luminosity is significantly smaller. 
Eventually, the lower bound on gluino is 1.55 TeV in the NUGM scenario.

The gray dashed lines show the branching ratio of 
loop induced two-body decay of gluino ,${\rm Br}(\tilde{g} \rightarrow g \tilde{\chi}_{1,2}^0)$. 
The rest of decay modes are three-body decays 
through the off-shell top squark dominantly or the off-shell bottom squark sub-dominantly. 
If the mass difference between the gluino and the neutralino is small, 
the loop induced two-body decay $\tilde{g} \rightarrow g \tilde{\chi}_{1,2}^0$ dominates the other decay modes. 
The two-body decay would not contribute to signal events 
since any signal region in this analysis requires at least three b-tagged jets. 
This fact relaxes the mass bound around the mass degenerate region.

\begin{figure}[!tb]
\centering
\hfill
\includegraphics[width=0.45\linewidth]{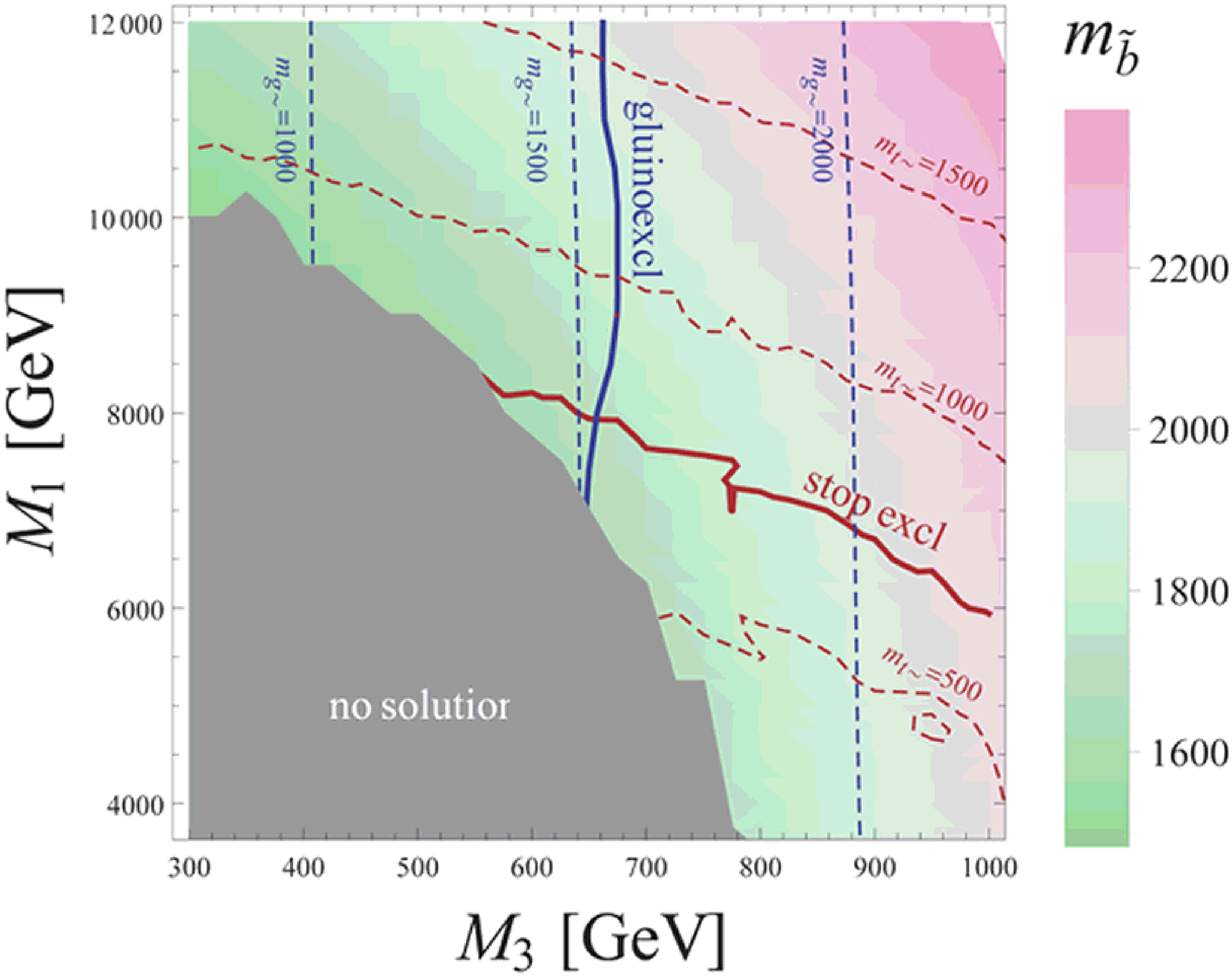}
\hfill 
\includegraphics[width=0.45\linewidth]{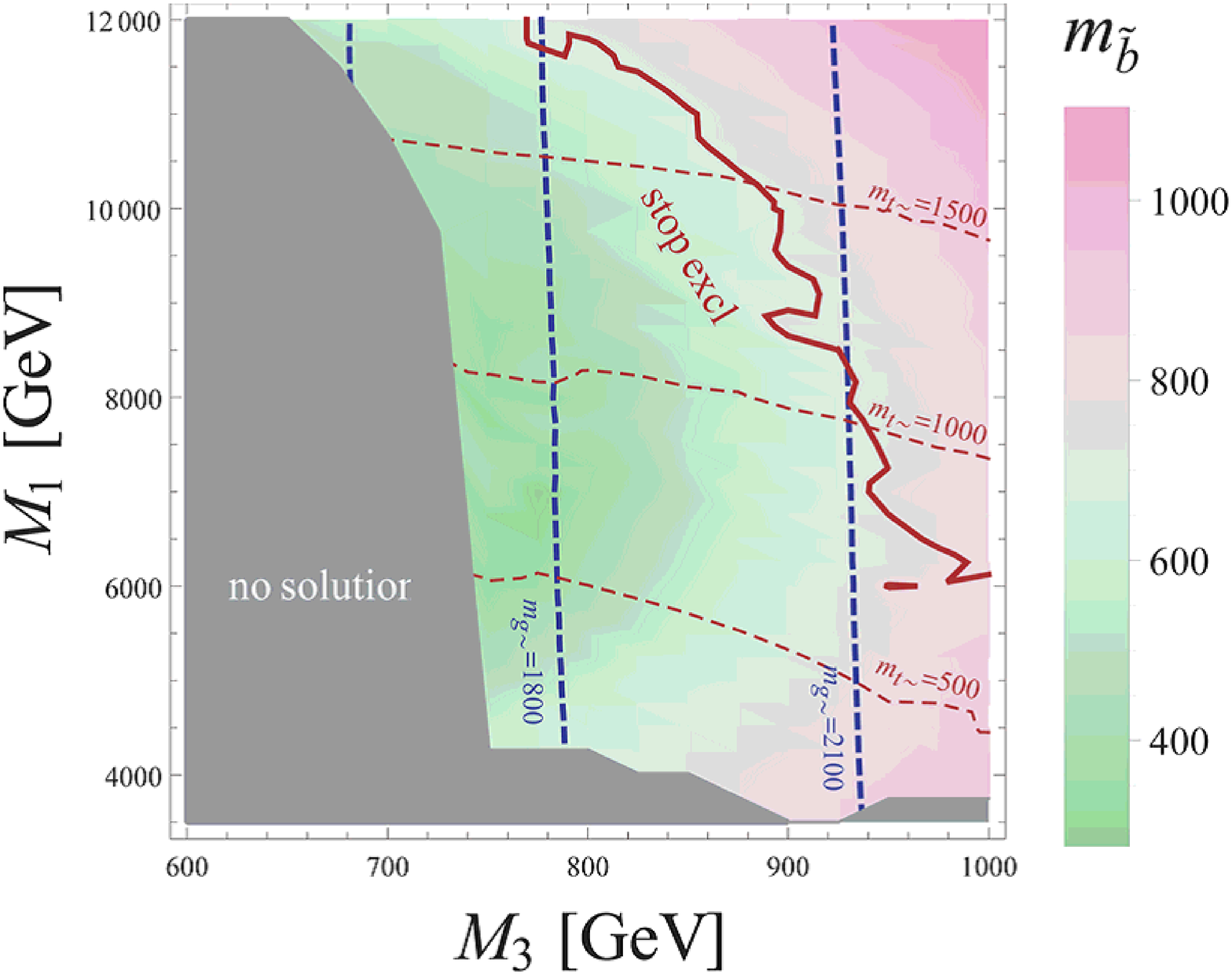} 
\hfill 
\caption{Exclusion limits on $M_3$-$M_1$ plane with $\mu = 150$ GeV and $\tan\beta=15$ (Left), $50$ (Right) 
from the top squark and gluino search are represented by the red and blue solid lines, respectively. 
The red and blue dashed lines show the top squark mass and the gluino mass. 
The suitable values for $M_2$ and $A_0$ are not found in the gray colored region. 
The background colors represent the bottom squark mass.   
All values are in the unit of GeV.
 }
\label{fig_M1M3}
\end{figure}

The projected bound on the $M_3$-$M_1$ plane is shown in the left panel of Fig.\ref{fig_M1M3} 
with $\mu = 150$ GeV. 
The suitable values for $M_2$ and $A_0$ are not found in the gray colored region 
since the top squark becomes tachyonic when $M_2$ is enough large to realize $\mu=150$ GeV 
and $m_h \simeq 126$ GeV.
The red and blue solid lines show the exclusion limits from top squark and gluino search 
at the LHC Run II, respectively.
The gluino can decay into on-shell top squark for $M_1\lesssim 12$ TeV, $M_3 \gtrsim 600$ GeV, 
but we can see that such top squark mass does not influence significantly to the exclusion limits on the gluino mass. We can see that the region $M_1 \lesssim 6.0$-$8.0$ TeV is excluded 
depending on $M_3$ according to the top squark search 
and $M_3 \gtrsim 650 $ GeV is required by the gluino search. 
Note that the exclusion limits will be relaxed as the $\mu $ parameter increases.

\subsection{Bottom squark search}

As mentioned in the previous section, the bottom squark production will give significant contributions 
to the signal events of the top squark searches 
if $\tan\beta$ is large. 
Such contribution would push up the exclusion limits. 

Behavior of bottom squark is quite similar to top squark 
since its branching ratio is almost identical with the top squark one.  
The fact that the bottom squark is almost right-handed as the top squark
leads the branching fractions of the bottom squark depend only on the mass difference against the higgsinos.  
A difference between the decays of the top and the bottom squark
is that the daughter higgsino is the nutralino or the chargino.
However, the higgisinos cannot be distinguished at the detector in our scenario
due to the degeneracy of the higgisinos.  
Therefore, the top squark and the bottom squark make almost identical signals at the LHC in the NUGM.
We estimate the contribution to the signal events from the bottom squark production 
by changing the normalization of the generated events of the top squark pair production.

Exclusion limit for $\tan\beta=50$ is shown in the right panel of the Fig.\ref{fig_M1M3}. 
The meanings for the lines and the colored regions are same as the left panel. 
The bottom squark becomes tachyonic in the region $M_3 \lesssim 750$ GeV, 
while the tau slepton becomes tachyonic when $M_1\lesssim 4.0$ TeV 
due to the large negative contribution from the tau Yukawa coupling in the RG effects. 
The theoretically excluded region is wider than the one in the small $\tan\beta$ case.
Especially it covers the parameter region within $m_{\tilde{g}} \lesssim 1.6$ TeV, which is excluded by the gluino search.

In this large $\tan \beta$ case, the bottom squark mass can be lighter than the top squark 
so that it significantly contributes to the top squark search.    
The red line show the exclusion limit from the $bb+E_T^{\rm miss}$ channel at the LHC Run II. 
The pure top squark contribution can only exclude $M_1 \lesssim 6.0-8.0$ TeV, 
while the limits can reach $M_1 \simeq 12$ TeV when $M_3 \simeq 800$ GeV 
due to the contributions of the bottom squark pair production.

\section{Conclusion}

In this paper, 
we investigate exclusion limits on the parameter space of the NUGM scenario 
where a natural SUSY spectrum is achieved due to a relatively heavy wino mass parameter.

We calculated the bound on the mass of top squark, 
which is almost right-handed 
and then it can decay into both $t \tilde{\chi}_{1,2}^0$ and $b \tilde{\chi}_1^{\pm}$. 
The top squark mass is roughly controlled by the bino mass parameter 
since the RG contributions from the gluino and the wino mass parameters are canceled each other 
in the light higgsino region. 
Thus the top squark searches at the LHC Run I and Run II can constrain
parameter region with the small bino mass parameter and the large gluino mass parameter.  
The top squark lighter than 700 GeV is excluded at $\mu \lesssim 150$ GeV, 
and lighter than 600 GeV is excluded at $\mu \lesssim 300$ GeV 
according to the result of the search for $bb + E_T^{\rm miss}$ at the LHC Run II. 
This limit already exceeds the one from the LHC Run 1 data. 
This lower bound corresponds to $M_1 \gtrsim 6.0$ TeV for $\mu \sim 150$ GeV and $M_1 \gtrsim 5.0$ TeV for $\mu \sim 300$ GeV, as can be seen in Fig. \ref{fig_stmu}. 
Note that there is no bound from the top squark search when $\mu \gtrsim 300$ GeV.

Gluino search is also good probes to the NUGM scenario due to the large production cross section. 
According to the LHC results, the 
parameter space with the small gluino mass and the large bino mass can be covered in our scenario. 
Note that the top squark is tachyonic in the parameter region 
where both bino and gluino masses are small.
The gluino mass less than 1.55 TeV is excluded by the ATLAS result at the LHC Run II 
when $\mu$ satisfies $\mu \lesssim 500$ GeV.  
We can conclude that the parameter region with $M_3 < 650$ GeV and $\mu \lesssim 500$ GeV is already excluded, as showed in Fig.\ref{fig_glmu}.

Bottom squark mass can be same or lighter than top squark mass
if $\tan\beta$ is so large that the bottom Yukawa coupling becomes sizable. 
Since the behavior of bottom squark at the collider experiment is quite similar to the one of top squark, 
the top squark search discussed above is also sensitive to the events generated by the bottom squarks. 
The exclusion limit on the parameter space of the NUGM is shown in Fig.\ref{fig_M1M3}. 
The wider region is prohibited theoretically compared with the small $\tan\beta$ case, 
in order to avoid the tachyonic bottom squarks or the tau slepton. 
The exclusion limit on the bino mass parameter reaches to $M_1 \simeq 12$ TeV for $M_3 \simeq 800$ GeV, 
and it reduces to 6.0 TeV as $M_3$ increases.

In this paper, we focus on the $bb + E_T^{\rm miss}$ channel to derive the  exclusion limits. 
The other channels will be important as more data at the LHC Run II is obtained. 
For instance, $tb + E_T^{\rm miss}$ channel could have same or more sensitivity for the top squark searches 
unlike the result of the LHC Run I,
since a critical mass of top squark would be enough large to produce energetic top quarks 
due to the larger mass difference against the higgsinos.  
Upcoming results at the LHC Run II will lead us to construct a true description behind the Standard Model.

 \afterpage{\clearpage}

\subsection*{Acknowledgements}
This work was supported in part by the Grant-in-Aid for Scientific Research No. 23104011 (Y.O.) from the Ministry of Education, 
Culture, Sports, Science and Technology (MEXT) in Japan. 

\appendix

\end{document}